\documentclass[prb,aps,twocolumn,showpacs,superscriptaddress,amsmath,amssymb]{revtex4}

\usepackage{graphicx}
\usepackage{psfrag}
\usepackage{epsfig}
\usepackage{color}
\usepackage[tight]{subfigure}
\definecolor{dred}{rgb}{0.7,0.0,0.0}
\usepackage{dcolumn}
\usepackage{bm}
\allowdisplaybreaks 

\begin{document}
\title{Emergent dimensional reduction of the spin sector \\ in a model for  narrow-band manganites}
\author{Shuhua Liang}
\affiliation{Department of Physics and Astronomy, The University of Tennessee, Knoxville, TN 37996, USA}
\affiliation{Materials Science and Technology Division, Oak Ridge National Laboratory, Oak Ridge, TN 32831, USA}

\author{Maria Daghofer}
\email{m.daghofer@ifw-dresden.de}
\affiliation{IFW Dresden, P.O. Box 27 01 16, D-01171 Dresden, Germany}

\author{Shuai Dong}
\affiliation{Department of Physics, Southeast University, Nanjing 211189, China}

\author{Cengiz \c{S}en}
\affiliation{Department of Physics and Astronomy, The University of Tennessee, Knoxville, TN 37996, USA}
\affiliation{Materials Science and Technology Division, Oak Ridge National Laboratory, Oak Ridge, TN 32831, USA}

\author{Elbio Dagotto}
\affiliation{Department of Physics and Astronomy, The University of Tennessee, Knoxville, TN 37996, USA}
\affiliation{Materials Science and Technology Division, Oak Ridge National Laboratory, Oak Ridge, TN 32831, USA}

\date{\today}

\begin{abstract}
The widely used Double-Exchange model for manganites is shown to support various
``striped'' phases at filling fractions $1/n$ ($n=3$, $4$,
$5 \dots$), in the previously unexplored regime of narrow bandwidth
and small Jahn-Teller coupling. Working in two dimensions,
our main result is that these 
stripes can be individually spin flipped without a physically relevant
change in the energy, i.e., we  find a large groundstate manifold
with nearly degenerate energies. The two-dimensional spin system thus 
displays an unexpected dynamically generated dimensional reduction
into decoupled one-dimensional stripes, even though the electronic
states remain two-dimensional.
Relations of our results 
with recent literature addressing compass models in
quantum computing are discussed.
\end{abstract}

\maketitle

\section{Introduction} \label{sec:intro}

The manganite family of Mn-oxide materials has attracted 
considerable attention mainly due to the
colossal magnetoresistance effect, where a magnetic field hugely 
enhances the electrical conductivity.~\cite{Dagotto:Prp}
More recently, multiferroicity has
been found in some members of this family, particularly when chemical substitution
reduces the bandwidth of the mobile electrons. Magnetic multiferroicity, where a
ferroelectric (FE) polarization is induced by the magnetic order, 
has lately been intensively investigated.~\cite{Cheong:Nm,Khomskii:Phy,Wang:Ap}
Furthermore, manganites provide a fertile ground to
study various ordered phases where magnetic, orbital, and charge
degrees of freedom interact and compete. As a result of the large
number of tendencies simultaneously active on comparable energy scales,
phases with very different physical properties can
nevertheless be very close in energy.~\cite{Hotta:Prl01,Hotta:Prl}

The competition of all these active degrees of freedom often leads 
to ``striped'' phases where  electron motion is
confined to one-dimensional (1D) subspaces.
In manganites at large hole doping, for example, electrons mainly occupy 
orbitals that point along either the $x$ or the $y$
direction.  Spins then order ferromagnetically along this
direction, because this favors the electronic kinetic energy via the
double-exchange (DE) mechanism. In the other direction, where the electronic motion is
suppressed, then DE is not active, and antiferromagnetic (AF) superexchange dominates. This
establishes AF spin correlations, leading to the ``spin striped'' $C$-type AF
phase.~\cite{PhysRevB.60.9506,PhysRevB.65.020407,PhysRevB.62.15096,PhysRevLett.82.1016,Dagotto:Prp}
Other examples involve the ferromagnetic (FM) zigzag chains, which are antiferromagnetically
coupled among themselves, that form the well-known 
CE-AF phase at half filling or the $E$-AF phase in the
undoped-limit.~\cite{Dagotto:Prp} The ground state of these phases
usually have collinear magnetic order, i.e., alignment between any two
spins is either perfectly FM or AF. Another aspect to note is that
while the electronic kinetic energy is 1D, the magnetic order is fully
two dimensional (2D), with AF order between the chains mediated by superexchange
(SE). In the absence of AF SE, a completely FM state with 2D electron
motion is energetically favorable.

Recently, some of us predicted the existence of a new 
multiferroic phase, dubbed 
\emph{spin-orthogonal stripe} (SOS) phase, located in the previously 
unexplored region of  quarter-hole-doping, small
Jahn-Teller electron-lattice coupling, and 
narrow $e_g$-electrons bandwidth.~\cite{Dong:Prl} In contrast to the
``striped'' phases mentioned above, the competition between FM DE and AF
SE stabilizes \emph{non-collinear} magnetic order on some bonds, where
the nearest-neighbor (NN) $t_{2g}$  spins are orthogonal to one another. 
 As in an analogous half-doped noncollinear
phase,~\cite{Giovannetti:Prl} this competition is expected to lead to
a relatively high multiferroic critical temperature ($\sim100$~K). 
These phases can be visualized as composed of ``thick'' stripes, where all magnetic
correlations within a stripe are collinear, i.e., AF or FM, while
adjacent stripes display non-collinear spins.

In the present publication, we report that the SOS phase described above, 
as well as similar
phases at dopings $x=1/n$  with integer $n>2$, have another unexpected
property, namely a very highly degenerate ground-state manifold. We
find that the spins of any collinear stripe can be rotated without a
significant change in energy, as long as the spins in adjacent stripes
remain orthogonal. Spins in
next-nearest-neighbor stripes can thus be at arbitrary angles
relative to each other, implying that there is no magnetic order in
the direction perpendicular to the stripes.
Analogous \emph{dimensional
reduction} effects on the magnetic order, where a higher-dimensional
(2D in our case) spin system decouples into lower-dimensional (1D)
subsystems, have been experimentally observed in other contexts such
as near 
a quantum-critical point.~\cite{Sebastian:2006p2468}
Three-dimensional spin-ice systems~\cite{Bramwell:Sci} 
also show macroscopic degeneracy, and they can decouple into 2D planes
when a magnetic field ``switches off'' some spins connecting the
planes.~\cite{PhysRevLett.97.257205,PhysRevB.68.064411}   

In contrast to spin ice, where  a local symmetry allows two states for
each tetrahedron, leading to a macroscopic ground-state degeneracy and
an extensive entropy proportional to the system size $N$, the near
degeneracy reported here is not quite macroscopic. The degeneracy
involves flipping or rotating whole stripes, and the number of stripes
in the 2D plane grows with the square root of the
system size. This indicates that the emergent degeneracy of the
ground state is intermediate between local and global. Similar
intermediate symmetry effects connected to dimensional reduction have been
extensively discussed in orbital-only models, as well as in the
context of quantum computation. An example is the so-called
``compass-model'' where the $x$ ($y$) components of the orbital
pseudo-spins are coupled Ising-like  in the $x$ ($y$) direction. This model 
was originally introduced to capture the
frustrated hoppings of the $e_g$ orbitals of manganites.~\cite{KK1982}
Its low-energy states are ordered along only one dimension,~\cite{Mishra:2004p2012,PhysRevB.71.195120}
implying a symmetry that, again, is intermediate between local and
global.~\cite{PhysRevB.71.195120,Nussinov:2009p2519}

Even though the compass model was originally suggested to model
the frustration inherent in degenerate $e_g$ orbitals in 2D, the
actual (partially frustrated) Hamiltonian describing these orbitals does not
exhibit such a high degeneracy. In fact, it has been shown that the
peculiar properties of the compass model are sensitive to even rather
slight modifications of the Hamiltonian and that a unique ground state with alternating
orbital order sets in easily when the model is modified towards a more
realistic description of 2D $e_g$ systems.~\cite{oles_mera}
In three dimensions, on the other hand, the model with a realistic $e_g$-orbital
structure does decouple into
\emph{planes},~\cite{PhysRevB.59.6795,Nussinov_orb_2004,PhysRevLett.105.146402}
as long as the magnetic order is fully FM. Once the magnetic degree of
freedom in included, however, the ground state turns out to show AF order along the
$z$ direction. This changes the relevant orbital Hamiltonian and as a consequence
three-dimensional spin-orbital order with a non-degenerate ground
state is stabilized. 
In addition to the $e_g$ orbital degrees of freedom in manganites,
heavier elements with a strong relativistic spin-orbit interaction were
discussed as a possible realization of compass models.~\cite{Jackeli:2009p2016} However, here
an isotropic Heisenberg term due to Hund's coupling 
would likely be present as well,~\cite{PhysRevLett.105.027204} again
inducing 2D order without $2^{\sqrt{N}}$ degeneracy.~\cite{Compass_Heisenberg} 
Finally, it was pointed out that the compass model describes properties
desired for fault-tolerant
qbits,~\cite{PhysRevB.71.195120,Doucot:2005p2466} and coupled
Josephson-junction arrays have been implemented for this
purpose.~\cite{Gladchenko:2008p2470}

The directional ordering described by the compass model is thus
potentially very interesting, but dedicated Josephson-junction arrays
appear so far the only systems showing such an effect. In the condensed-matter systems
conjectured to display this physics based on particular features of their
low-energy Hamiltonians, additional - even
rather small~\cite{Compass_Heisenberg} - perturbations, as often
present in realistic descriptions of materials, have thus far been found to lift the high
degeneracy. It has been suggested that the \emph{opposite} route
might work, i.e. that a more complex Hamiltonian, which  does not
itself have the appropriate symmetry intermediate between local
and global, might still support a ground-state manifold that
\emph{has} such
symmetries.\cite{Nussinov_IJPB_2006,Nussinov:2009p2519,Nussinov2009977} 
As mentioned above, some experimental evidence of such behavior
exists,~\cite{Sebastian:2006p2468} and a similar effect is known in
the case of spin-ice.~\cite{PhysRevLett.97.257205,PhysRevB.68.064411}
However, we are not aware of any model where ``compass-like'' behavior
has been shown to emerge effectively for the ground-state manifold. We
present here an unbiased numerical study showing this to happen in a
Hamiltonian realistic for manganites, in a particular region of parameter space. 

Section~\ref{sec:model} contains the Model, which describes not
only the $e_g$ orbital system at the origin of the compass-model, but
also includes the spin and even the coupling to lattice distortions. In
Sec.~\ref{sec:methods}, the numerical techniques are presented. In
Sec.~\ref{sec:results}, we show that the 2D spin system effectively
decomposes into uncoupled 1D stripes, while the electronic kinetic energy
remains fully 2D. We also discuss orbital occupation and the relation
between the magnetic degeneracy and dispersionless electronic states.

\section{Model} \label{sec:model}

The Hamiltonian considered here has been extensively studied in the
past decade, and it has been shown to be very helpful to understand the complex behavior of 
manganites. In particular, the model reproduces the large variety of phases
observed in manganites, e.g. $A$-type AF, $C$-type AF, FM, CE-AF, or
$E$-AF, and also the colossal-magnetoresistance regime.~\cite{Dagotto:Prp} 
The Hamiltonian is given by
\begin{align}\label{eq:ham}
\nonumber H&=-\sum_{\langle ij \rangle\parallel
  x/y}^{\alpha,\beta}t_{x/y}^{\alpha\beta}(\Omega_{ij}c_{i\alpha}^{\dagger}c^{\phantom{\dagger}}_{j\beta}+H.c.)
+J_{\rm AF}\sum_{<ij>}\vec{S}_{i}\cdot\vec{S}_{j}\\
\nonumber &+\lambda\sum_{i}(-Q_{1i}n_{i}+Q_{2i}\tau_{i}^{x}+Q_{3i}\tau_{i}^z)\\
&+\frac{1}{2}\sum_{i}(2Q_{1i}^2+Q_{2i}^2+Q_{3i}^2).
\end{align}
The first term gives the kinetic energy of electrons in the two $e_g$
orbitals, containing the electronic hopping on NN bonds $\langle$$ij$$\rangle$ along the $x$ and $y$
directions (we study a 2D lattice for simplicity). 
The operator $c_{i\alpha}^{\dagger}$ ($c_{i\alpha}^{\phantom{\dagger}}$)
creates (annihilates) an electron at site $i$ and at orbital $\alpha$,
where the orbital indices $\alpha$ and $\beta$ run over the $d_{x^2-y^2}$ and $d_{3r^2-z^2}$ orbitals
of the Mn ions. The orbital- and direction-dependent hopping
parameters are given 
by 
\begin{equation}
t_{x}^{\alpha\beta}=\left(\begin{matrix}\frac{3}{4}&
    -\frac{\sqrt{3}}{4}\\ -\frac{\sqrt{3}}{4}&\frac{1}{4}\end{matrix}\right)t_0,\quad
t_{y}^{\alpha\beta}=\left(\begin{matrix}\frac{3}{4}&
    +\frac{\sqrt{3}}{4}\\ +\frac{\sqrt{3}}{4}&\frac{1}{4}\end{matrix}\right)t_0,
\end{equation}
where the interorbital hoppings are negative (positive) on bonds along
the the $x$ ($y$) direction, and
$t_{0}=0.2$-$0.5$~eV defines the energy unit.~\cite{Dagotto:Prp} The Hund's coupling, which links the
itinerant electrons to localized $t_{2g}$ spins, is here taken to be
infinite for simplicity, which implies that the $e_g$ electrons' spin is always
parallel to the local $t_{2g}$ spin. Neither Hund's rule coupling nor
the electron spin thus appear explicitly in the Hamiltonian. This approach leads to a
modification of the bare hopping that is captured by a (complex) Berry phase factor
\begin{equation}\label{eq:berry}
\Omega_{ij}=\cos \frac{\theta_{i}}{2} \cos \frac{\theta_{j}}{2}
+\sin \frac{\theta_{i}}{2} \sin \frac{\theta_{j}}{2}\textrm{e}^{-i(\phi_{i}-\phi_{j})},
\end{equation}
which depends on the angles $\theta_{i}$, $\phi_i$ and $\theta_j$, $\phi_j$ that define the classical
localized spins at sites $i$ and $j$.\cite{PhysRevB.54.R6819} Between sites with parallel (antiparallel)
$t_{2g}$ spins, the Berry phase factor becomes one (zero),
implying that the kinetic energy favors parallel spins due to the
DE mechanism. Between non-collinear spins, i.e., spins with a relative
angle between 0 and 180 degrees, its absolute value is between 0 and
1, and it can be negative or even complex.

The second term of Eq.~(\ref{eq:ham}) describes the direct AF SE coupling between NN
$t_{2g}$ spins. The third terms represents the coupling of
$e_g$ electrons with the lattice, via the Jahn-Teller (JT) 
($Q_2$ and $Q_3$) and breathing ($Q_1$) modes. $\lambda$ is a dimensionless lattice-electron coupling
coefficient. Only the $x$-$y$ plane distortions are considered here,
and if the overall lattice shape (square) can be assumed not to
change, $Q_{1}$ can be set to $-\sqrt{2}Q_{3}$. The
lattice normal modes $Q_2$ and $Q_3$ are related to shifts 
$\delta^{x}$, $\delta^{y}$ ($\delta^{z}=0$) of the coordinates of the
six oxygens surrounding each manganese via $Q_{2}=(\delta^{x}-\delta^{y})/\sqrt{2}$ and
$Q_{3}=-(\delta^{x}+\delta^{y})/\sqrt{6}$. $n_{i\alpha} =
c^{\dagger}_{i\alpha}c^{\phantom{\dagger}}_{i\alpha}$ and $n_i =
n_{i,x^2-y^2}+n_{i,3z^2-r^2}$ are density operators; while
\begin{align}
\tau_{i}^{z}
&= (n_{i,x^2-y^2}-n_{i,3z^2-r^2})/2, \\ 
\tau_{i}^{x} &=
(c^{\dagger}_{i,x^2-y^2}c^{\phantom{\dagger}}_{i,3z^2-r^2}+c^{\dagger}_{i,3z^2-r^2}c^{\phantom{\dagger}}_{i,x^2-y^2})/2,
\end{align} 
denote the orbital pseudospin operators, 
similar to the Pauli matrices for spins.~\cite{Dagotto:Prp} The
lattice distortions are thus coupled to the orbital degree of
freedom. 
Undoped manganites correspond to a filling of one electron per Mn, while
doping $x$ and filling $n$ are related via $n=1-x$.

\section{Methods} \label{sec:methods}

Markov-chain Monte Carlo (MCMC), zero-$T$ optimization,
and variational methods were used to study the ground-state and
low-temperature properties of the Hamiltonian Eq.~(\ref{eq:ham}). This
Hamiltonian couples non-interacting electrons to  the classical
variables ${\vec
  q}_{i}=\{\theta^{\phantom{x}}_{i},\phi^{\phantom{x}}_{i},
\delta^{x}_{i},\delta^{y}_{i}\}$ describing the localized spins and
lattice distortions. For any set of ${\vec q}$'s, Eq.~(\ref{eq:ham})
is diagonalized using standard library routines, and the free energy
of the electronic system can be easily evaluated via the usual 
equations from statistical physics. The energy is then used in a
conventional MCMC procedure to determine whether a new set of ${\vec 
  q}_{i}$ should be accepted, as detailed in various
publications.~\cite{Dagotto:Prp} Since care must be taken that the
results are independent from the initial configuration and that thermal
equilibrium has been reached, such MCMC simulations are quite CPU-time
consuming. Their huge advantage is that they are \emph{unbiased},
meaning that, for long enough run time, they will converge to the true
relevant state of the cluster under study, regardless of the initial
state used. We employed this method on $6\times 6$ lattices
to obtain the phase diagram  in the
$\lambda$-$J_{\rm AF}$ plane, and on $12\times 12$ lattices 
for a selected number of points.

The MCMC is complemented by the zero-$T$
optimization method where the ${\vec q}$'s are optimized to reach the
lowest possible energy by employing the derivatives
$\frac{\partial{H}(q)}{\partial{q}}$, as detailed in Ref.~\onlinecite{Yu:Prb09}. This
optimization method is particularly useful around a local energy
minimum, where it reaches higher precision than the MCMC and is, thus,
efficiently combined with MCMC, which is better at finding the \emph{global}
minimum. Finally, we also complement these studies by a variational comparison of
the energies of fixed configurations of classical spins and
lattice distortions. While the variational approach is not unbiased,
because only chosen configurations were combined, it is valuable,
because far larger lattices can be reached; it was employed, e.g., to
verify that various SOS$_{1/3}$ configurations indeed are practically
degenerate. In addition to $L\times L$ squares, this approach was
also employed on $\sqrt{2}L_1\times\sqrt{2}L_2$ rectangles. Periodic
boundary conditions were used in MCMC and optimization, and for
variational energy comparison. To calculate selected observables,
such as the spin-structure factor and the density of states,  we
additionally used twisted boundary conditions leading to a denser $k$
mesh, as described in, e.g., Ref.~\onlinecite{Salafranca:Prb09_tbc}.

\section{Results} \label{sec:results}

Motivated by the recent prediction of the $x=1/4$ SOS
phase,~\cite{Dong:Prl} and of a similar phase at half-doping\cite{Giovannetti:Prl}
we have investigated other fillings
$x=1/3$, $1/5$, $\dots$, 
focusing here on parameters relevant to narrow-band manganites.
The discussion below mainly focuses on results for $x=1/3$, where the
SOS phase has narrower stripes than for smaller $x$ and can thus be
investigated on smaller clusters. A variational energy comparison and
some MCMC studies were also performed for $n\geq 5$, leading to
analogous results.

\subsection{Phase diagram and highly degenerate ground state manifold} \label{sec:phases}

\begin{figure}
\subfiguretopcaptrue
\subfigure{\includegraphics[height=0.22\textheight]{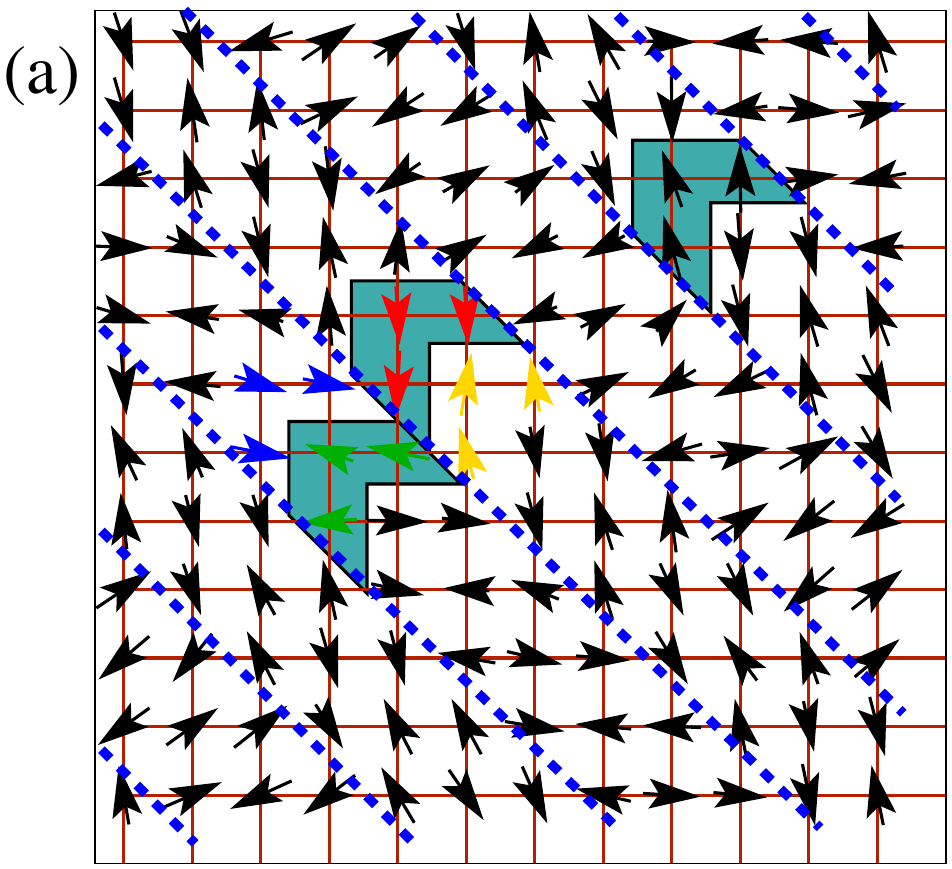}\label{fig:SOS_spins}}
\subfigure{\includegraphics[height=0.22\textheight]{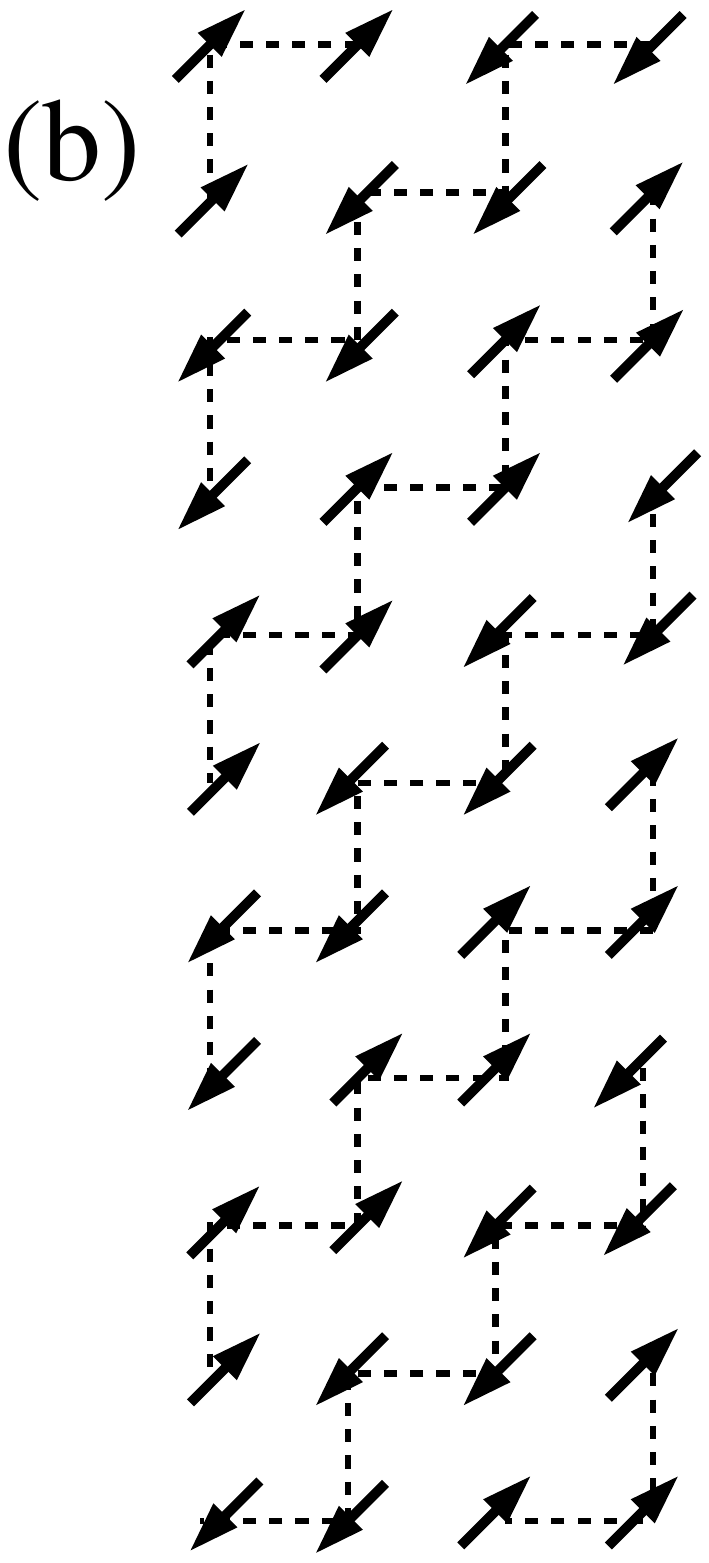}\label{fig:EAF_spins}}\\
\caption{(color online) 
(a) Monte-Carlo snapshot for $J_{\textrm{AF}}=0.19t_0$, $\lambda =0$, and
$\beta t_0 = 1200$, in the regime of the SOS$_{1/3}$ phase. It can be
visualized as composed of domains of the $E$-AF phase illustrated in
the cartoon (b). NN domains (``stripes'') are separated by dashed lines in (a), they
have spins at right angles. In (b), dashed lines indicate the zig-zag FM chains of 
the $E$-AF phase; in (a)  shading (color) indicates a few of the short
segments of the $E$-phase zigzag-chains that survive in the
SOS$_{1/3}$ phase, later called ``arrows''.\label{fig:sos_e}}
\end{figure}

\begin{figure}
\includegraphics[width=0.30\textwidth,clip]{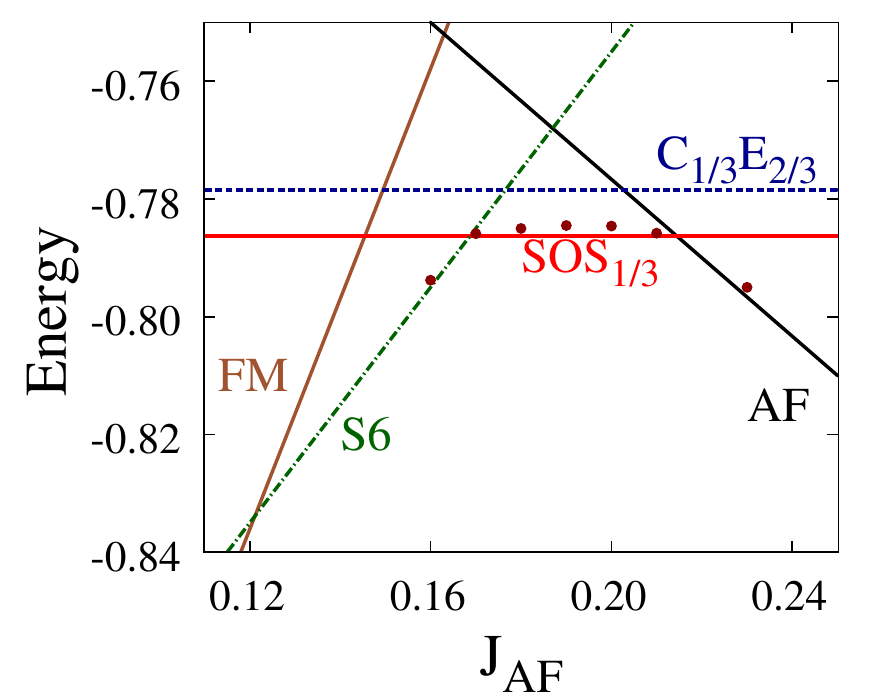}
\caption{(color online) 
Zero temperature ($T=0$) energies (per site) of several magnetic phases, at doping $x=1/3$ 
and $\lambda=0$, vs. $J_{\rm AF}$. Dots represent MCMC results at
$T=0.002$ on a $6\times 6$ lattice that closely follow the variational results. 
S6 denotes a spiral phase with a period of 6 lattice
spacings,~\cite{Dong:Prb08.2} while a $12\times 12$ lattices shows one
with period 12; this phase thus converges to FM order 
with increasing lattice sizes.~\cite{Riera:Prl} \label{Variational} 
}
\end{figure}

\begin{figure}
\includegraphics[width=0.49\textwidth]{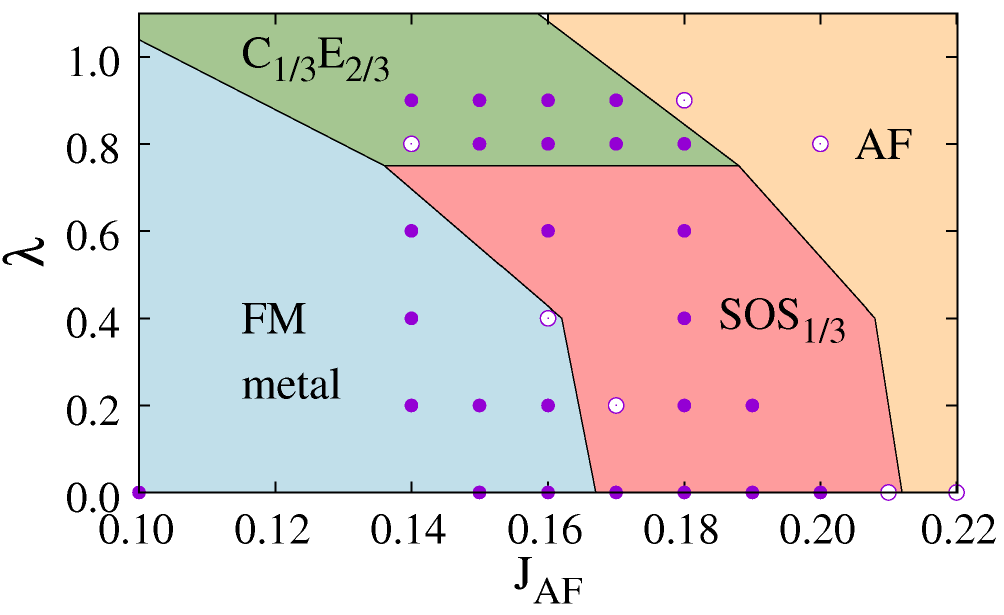}
\vspace{-0.2cm}
\caption{(color online) 
Zero-$T$ phase diagram of model Eq.(1) at $x$=1/3, varying $\lambda$ and $J_{\rm AF}$.
Variational technique results are shown with shading (colors). The
FM DE metallic phase dominates; large $J_{\rm AF}$ stabilizes AF phases. 
The ``C$_{1/3}$E$_{2/3}$'' phase is a
variant~\cite{Dong:Prl,Hotta:Prl} of the CE phase at half filling.~\cite{Dagotto:Prp} 
The SOS$_{1/3}$ phase at intermediate $J_{\rm AF}$ and small $\lambda$ is
analogous to the $x$=1/4 SOS$_{1/4}$ phase,~\cite{Dong:Prl} and is
\emph{highly degenerate}. Full dots show where MCMC (for $6\times 6$
sites) confirmed the results;
open dots are MCMC results that remained inconclusive due to
metastabilities caused by phase competition. For details see Ref.~\onlinecite{Dong:Prl}.
 \label{fig:phasediagram}}
\vspace{-0.3cm}
\end{figure}

Figure~\ref{fig:sos_e}(a) shows a snapshot obtained in a MCMC run for
$J_{\textrm{AF}}=0.19t_0$ and $\lambda =0$, where the spins happened
to lie almost within the $x$-$y$ plane. It illustrates the
SOS-phase expected for a filling of $1/3$, actually just one of its
realizations, see the discussion below.  
As discussed for $x=1/4$,~\cite{Dong:Prl} the SOS phase
consists of domains of the $E$-AF phase of undoped
manganites [illustrated in Fig.~\ref{fig:sos_e}(b)], with spins rotated by 90$^\circ$ between neighboring
domains. Each domain can be visualized as one ``stripe'', see
Fig.~\ref{fig:sos_e}(a), and the spins are then collinear within each of the
stripes (regions between a pair of dashed lines), but between stripes they are
orthogonal to each other. 
In Fig.~\ref{Variational}, the groundstate energy of the SOS$_{1/3}$
phase is compared to that of various other phases in the absence of
electron-phonon coupling ($\lambda=0$), and it is clear that the
SOS$_{1/3}$ phase  has the lowest energy for a range of  $J_{\rm
  AF}$. This is corroborated by 
data points obtained with unbiased MCMC, which closely follow the
variational energies, indicating that the true ground state has been
found. 

The phase diagram including electron-phonon coupling $\lambda$ in
addition to $J_{\rm AF}$ is given in Fig.~\ref{fig:phasediagram}, and it
shows that the SOS$_{1/3}$ phase remains a stable ground state for
$\lambda \lesssim 0.7$. As it may be expected,  DE drives a FM metallic
state at smaller $J_{\textrm{AF}}$, while large $J_{\rm AF}$
stabilizes a fully $G$-type AF phase. At large $\lambda$ and
intermediate $J_{\textrm{AF}}$, the SOS$_{1/3}$ phase is replaced by a
variant of the exotic but well studied~\cite{Hotta:Prl,Dagotto:Prp} CE-phase.

\begin{figure}
\subfigure{\includegraphics[width=0.15\textwidth]{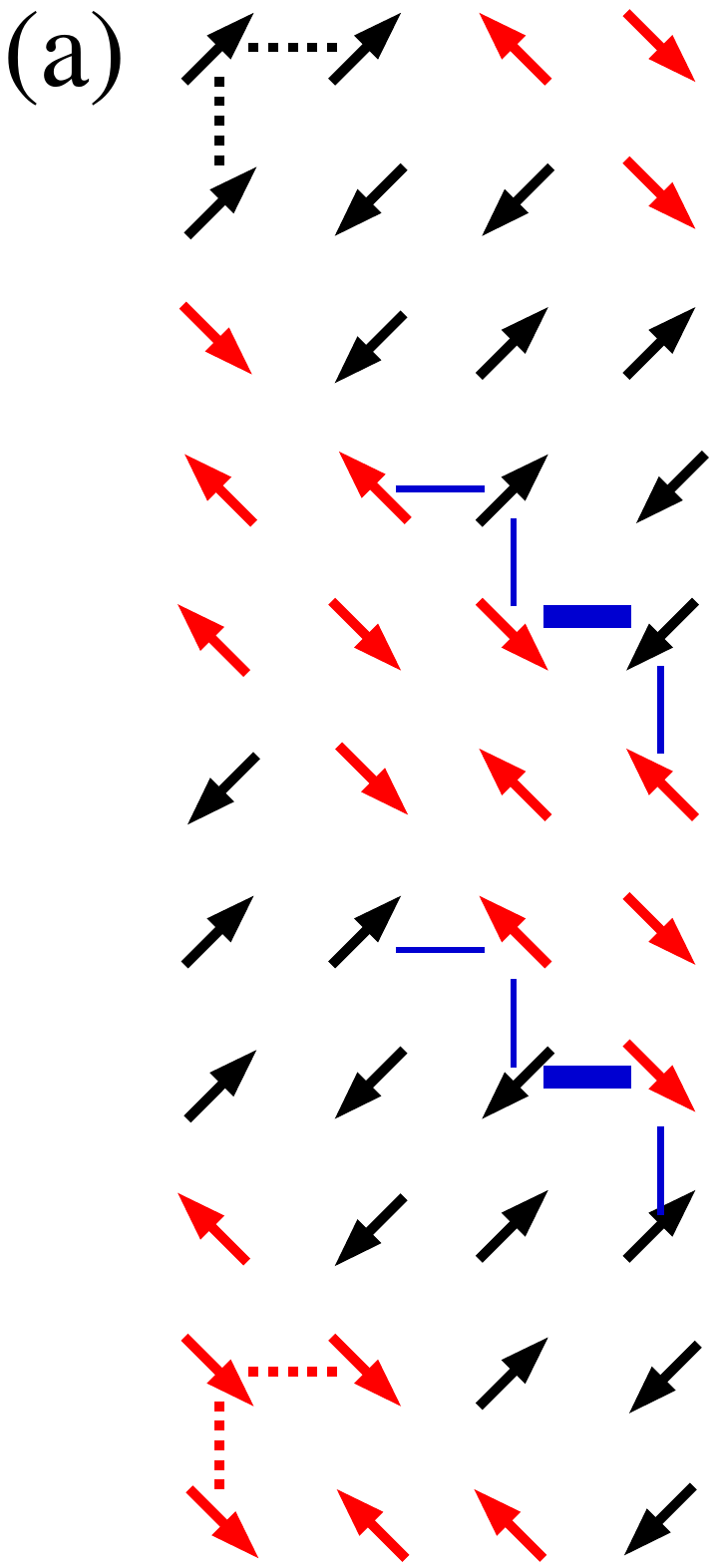}}
\hfill
\subfigure{\includegraphics[width=0.15\textwidth]{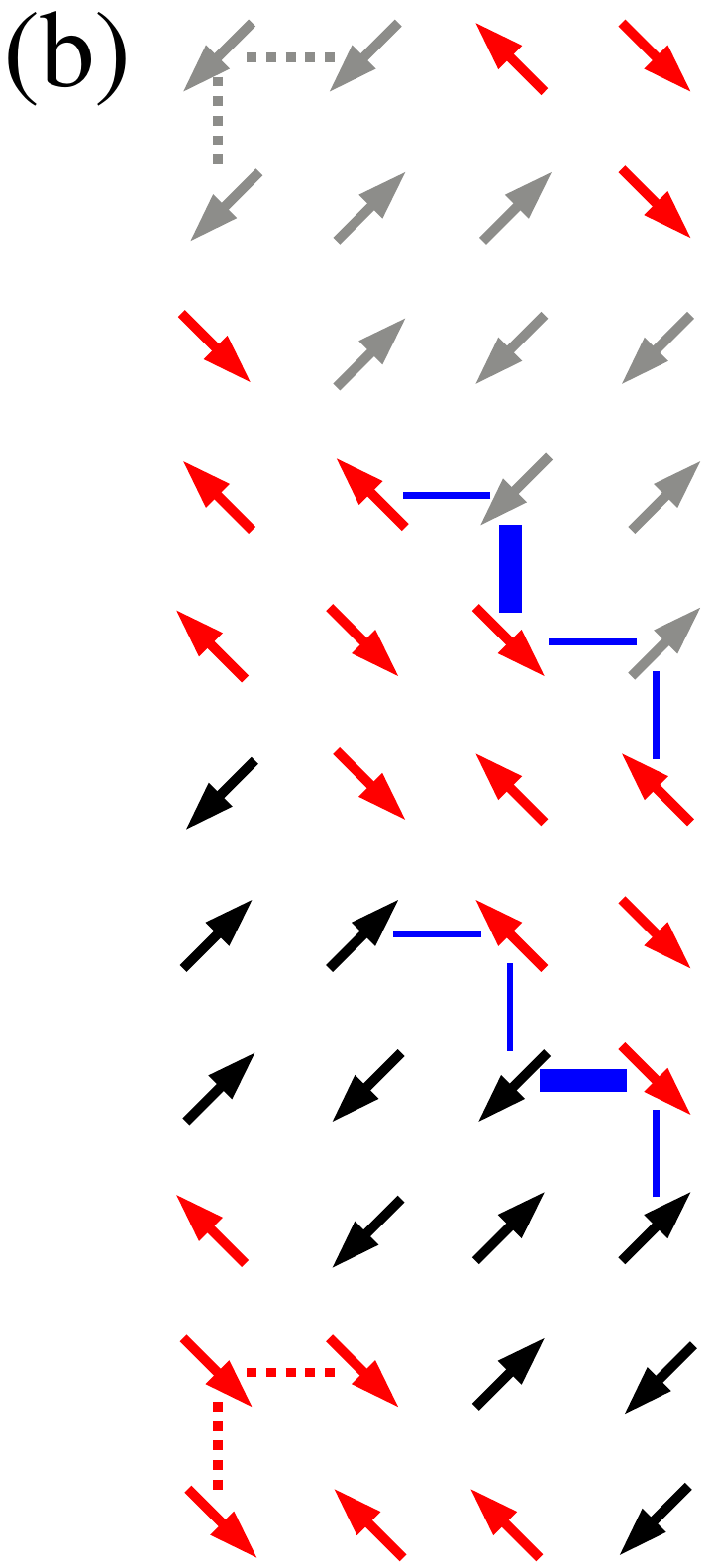}}
\hfill
\subfigure{\includegraphics[width=0.15\textwidth]{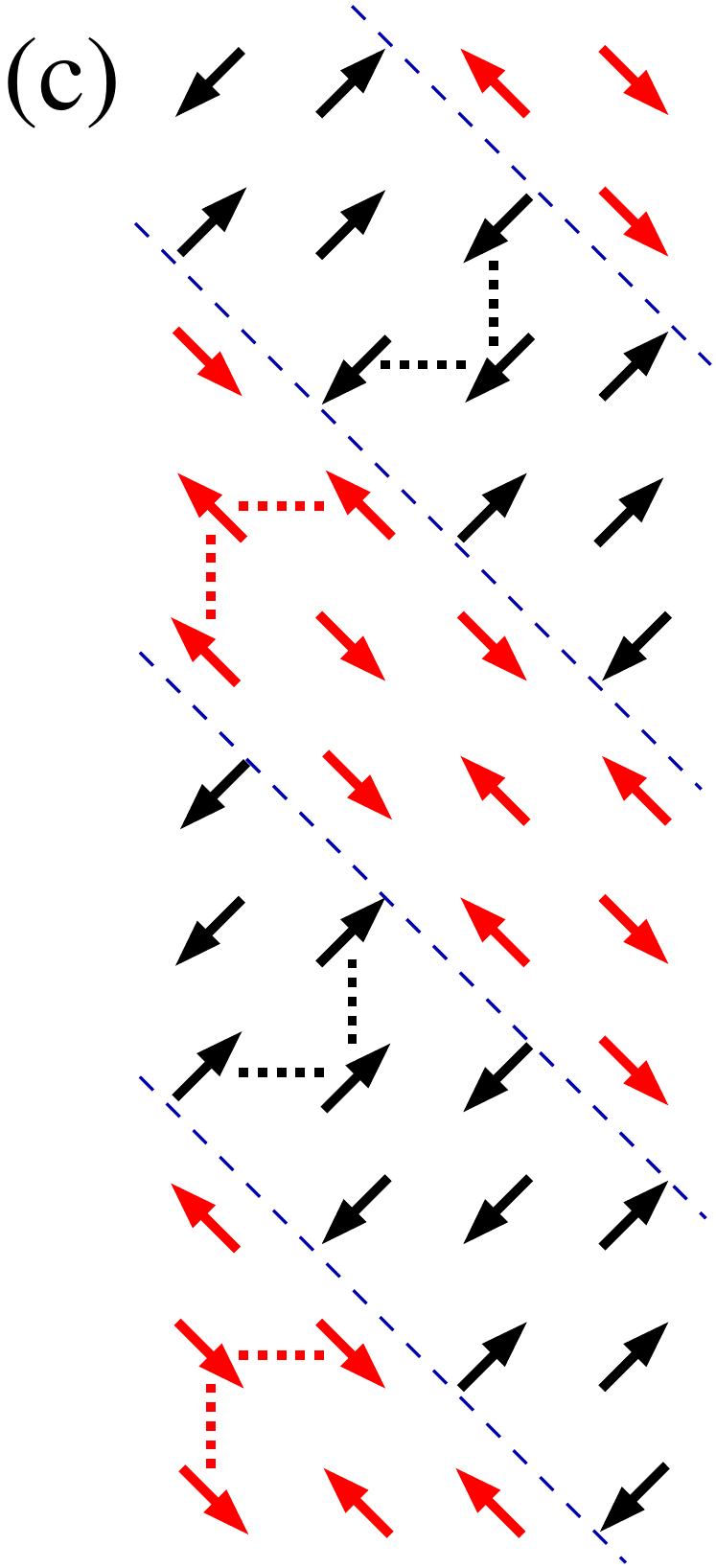}}\\
\caption{(color online) Cartoon for the spin patterns for two
  quasi-degenerate states of the SOS$_{1/3}$ phase (a-b). Lines
  between domains with orthogonal spins illustrate the periodic
  pattern of Berry phases: Thin lines denote positive sign, thick lines
  negative. The pattern in (b) is reached by flipping all spins
  in the top black stripe of (a), which entails changes in the signs of
  the Berry phases, but induces an energy difference of
  only $10^{-5}t_0$. 
  The dotted lines connecting three ferromagnetically aligned spins indicate `arrows' of the
  $E$-phase, see Fig.~\ref{fig:sos_e}.  
  In the pattern in (c), `arrows' in adjacent domains point in
  opposite directions. The energy per site in this phase is
  $10^{-2}t_0$ higher than that of the SOS$_{1/3}$ phase; flipping spins of one stripe
  induces energy differences of $\approx 5\cdot 10^{-4}t_0$, i.e., at
  least 50 times larger than between (a) and (b).\label{SOS3}}
\end{figure}

However, our main result is that
our calculations 
have revealed that this ground state is {\it not unique}: if all spins
within one stripe are flipped by 180$^\circ$ (i.e. inverted), the total energy remains
nearly unchanged. This is illustrated in Fig.~\ref{SOS3}(a-b), where
two almost degenerate variants of the SOS$_{1/3}$ phase are
illustrated. Taking into account non-coplanar spin
configurations as well, the spins in each stripe can in fact
be rotated by any angle, as long as the spins in adjacent stripes remain
orthogonal. 

On square clusters, the difference in energy per site
before and after a stripe spin-flip is merely $\sim 10^{-5}t_0$ for
flipping one stripe of a $12\times 12$ lattice. This is three
orders of magnitude smaller than the energy differences with other states 
shown in Fig.~\ref{Variational}; similar conclusions were reached
using tilted rectangles. 
Finite electron-phonon coupling $\lambda >0$ as well as finite Hund's
rule coupling $J_{\textrm{Hund}}<\infty$ leave the
near degeneracy intact. Even rather small Hund's coupling 
$6t_0\approx 1.2$~eV increases the energy split only to $\approx 5\cdot10^{-5}
t_0$, still more than two orders of magnitude smaller than the energy
differences to other phases ($\lambda\gtrsim 0.7t_0$ eventually drives
a transition to the C$_{1/3}$E$_{2/3}$ state, see
Fig.~\ref{fig:phasediagram}.) 

\subsection{Magnetic, orbital and charge patterns}\label{sec:spinorb}

\begin{figure}
\includegraphics[width=0.47\textwidth,trim = 0 0 0 160,clip]{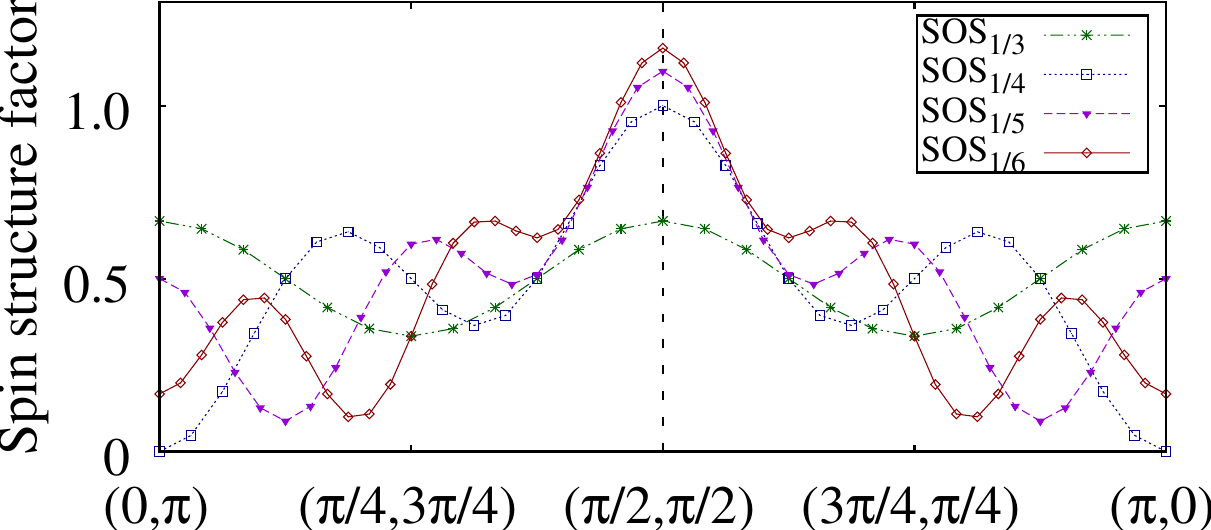}
\caption{(color online)  Spin-structure factor $S({\bf k})$ along the
  line $(\pi, 0)$ to $(0,\pi)$, calculated for the SOS$_{1/3}$
  phase by averaging over all $2^{16}$ degenerate realizations with stripes along
  one direction of a $24\sqrt{2}\times\sqrt{2}$ cluster ($\lambda$=0). $S({\bf
    k})\approx 0$  for all ${\bf k}$ except the line running from
  $(\pi, 0)$ to $(0,\pi)$. Results at $x$ = 1/4, 1/5, and 1/6 are also shown.
\label{fig:sk_sos}}
\end{figure}

Effectively, the 2D system decouples into 1D stripes, whose
direction can be rotated independently from the other stripes as long
as spins in NN stripes are at right angles. The relative orientation
of stripes at larger distance is thus arbitrary and the spin structure factor 
$S({\bf k})$ is finite along a whole \emph{line} in
momentum space, see Fig.~\ref{fig:sk_sos}(b), similar to results for the
compass model.~\cite{Compass_Heisenberg} The $S({\bf k
})$ modulation is due to the width of a double stripe.

Since spins on all bonds between stripes are orthogonal, the absolute
value of the hopping connecting the stripes is the same in all
realizations of the SOS$_{1/3}$ phase, with $|\Omega_{ij}|=1/\sqrt{2}$
obtained from (\ref{eq:berry}). However, having the same
$|\Omega_{ij}|$ is not enough to establish such a degeneracy, as the complex phase of the
Berry phase in general cannot be neglected. In the ``flux''
phase,~\cite{flux} NN spins are always orthogonal and flipping a spin
would not change this; but there is nevertheless a unique ground state
stabilized by the Berry phase. Similarly, flipping a stripe of the
``SOS$_{1/2}$'' phase at half doping~\cite{Giovannetti:Prl} preserves the
absolute value of all hoppings, yet costs far more energy
than flipping a stripe of the SOS$_{1/3}$ phase. In order to show that
the groundstate degeneracy is not due solely to having orthogonal
spins, it is illustrative to analyze the ``modified'' SOS$_{1/3}$ phase
depicted in Fig.~\ref{SOS3}(c). Like the actual SOS$_{1/3}$ phase, this
phase is made up of domains of the $E$-AF phase that are orthogonal
to each other. The only difference is that the ``arrows'' formed by the FM
spins point in opposite directions in adjacent stripes in this
modified phase. Even though the spins along the boundaries between
stripes are the same in both phases, the modified phase has a higher
energy per site by $10^{-2}t_0$, which is larger that the energy
difference to the C$_{1/3}$E$_{2/3}$ phase, see
Fig.~\ref{Variational}, indicating that the internal composition of
the stripes matters as much as their boundaries. Moreover, the
modified phase does not show such a near perfect degeneracy, as flipping a
stripe on a $12\times 12$ lattice costs  $\approx 5\cdot 10^{-4}t_0$,
and while this is not a large energy difference, it is at least 50
times as much as for the SOS$_{1/3}$ phase. 

\begin{figure}
\subfigure{\includegraphics[width=0.21\textwidth,angle=-90]{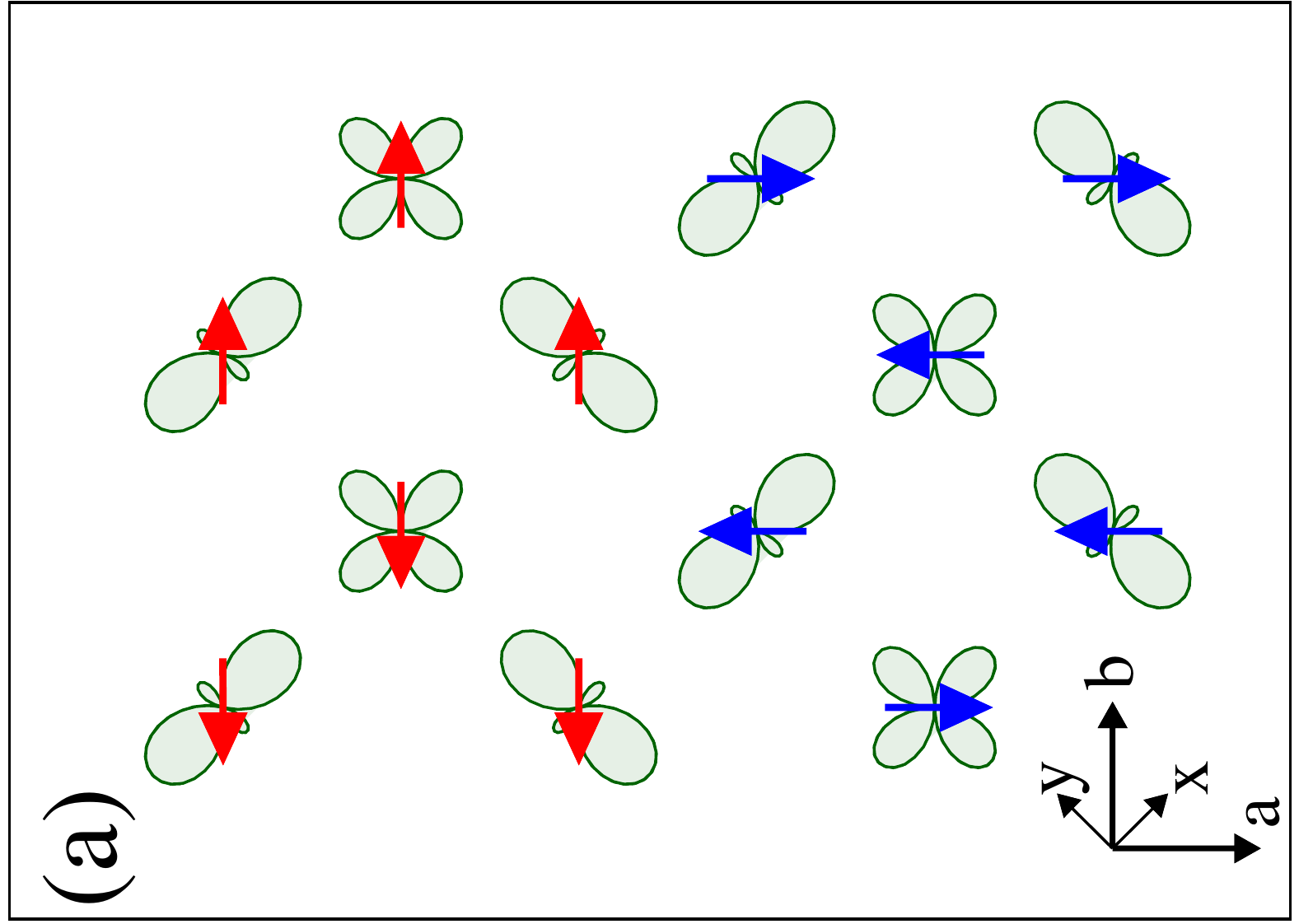}\label{CO3}}
\hspace*{0.05\textwidth}
\subfigure{\includegraphics[width=0.21\textwidth,angle=-90]{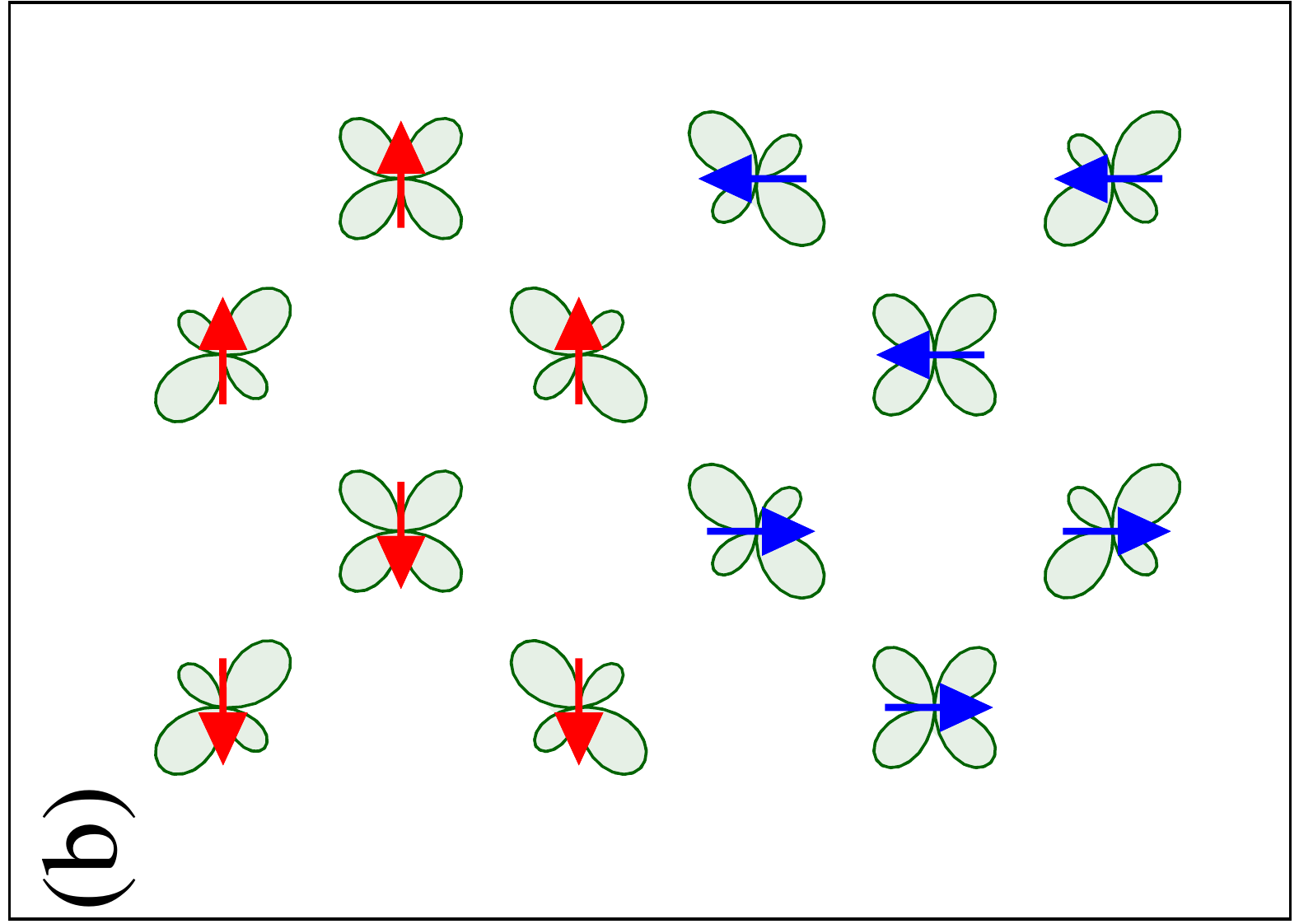}\label{CO3_wrong}}\\
\caption{(color online) 
(a) Zero-$T$ spin and orbital ordering of the SOS$_{1/3}$ phase, the
ground state for $0.17\lesssim J_{\rm AF}\lesssim 0.21$ ($\lambda=0$), see
Fig.~\ref{fig:phasediagram}. For each site, the linear
combination $|\alpha\rangle = \cos \alpha |3z^2-r^2\rangle + \sin \alpha
|x^2-y^2\rangle$ is depicted that has the highest density, i.e., for the value of
$\alpha$ maximizing $n_{i,\alpha}$. In (b), the orbital with maximal
density is shown for the modified SOS$_{1/3}$ phase with the spin
pattern given in Fig.~\ref{SOS3}(c). While the depicted orbital is the
one with the highest density, these cartoons do not fully describe the
orbital states, as some density is also found in the orthogonal
orbitals, see text.
\label{fig:orbitals}}
\end{figure}

The dominant orbital occupancy and its relation to the spin pattern is
shown in Fig.~\ref{CO3}. The building block of the SOS$_{1/3}$
phase is an arrow made of three FM spins, with a `center' and two `wings'
that point in the $x$ and $y$ directions. For each site, one can
calculate the dominant linear combination of the two $e_g$ orbitals,
i.e., the orbital with the highest density.  At the wings of the
arrows, these are the directional orbitals pointing to the center,
because they can maximize the kinetic energy along the 
FM bond,\cite{Dagotto:Prp} see Fig.~\ref{CO3}; they are half filled. On the central site, the
$x^2-y^2$ orbital dominates due to its large overlap with
adjacent sites.  However, it turns out that the electronic 
configuration of the SOS$_{1/3}$ phase cannot be fully characterized in terms of
singly occupied orbitals, which is reminiscent of the edge sites in the
CE-phase.~\cite{vandenBrink:1999p2416} Even though the depicted orbitals have the highest
density, some electronic weight is also found in the orthogonal
states  and Fig.~\ref{CO3} only partially describes
the orbital state. In the center, the explicit
occupancies are thus $n_{x^2-y^2} = 0.4$ and $n_{3z^2-r^2} = 0.28$,
and in the `wings' along $x$/$y$, we find $n_{y^2-z^2/z^2-x^2}=0.24$
in addition to $n_{3x^2-r^2/3y^2-r^2}=0.5$.  The total density
is the sum of the two orbital densities on each site, and shows only
weak stripe modulation, with $n=0.68$ in the center and $n=0.66$ in
the wings. 

The relatively large density
$n_{y^2-z^2}=0.24$ in a `wing' pointing along $x$ may seem surprising, as the $y^2-z^2$ orbital cannot hop
along the FM bond directed along $x$ to the center.\cite{Dagotto:Prp} However, it can hop with an
only slightly reduced amplitude $|\Omega_{ij}|=1/\sqrt{2}$ to an
adjacent stripe. This process becomes even more important in the
modified SOS$_{1/3}$ phase of Fig.~\ref{SOS3}(c), whose dominant
orbitals are shown in Fig.~\ref{CO3_wrong}. In this case, the
$1/\sqrt{2}$ inter-stripe hopping connects two directional $3x^2-r^2$
(or $3y^2-r^2$) orbitals with a large overlap. We will discuss in the
next Section~\ref{sec:ekin} why the hoppings between stripes cannot
be neglected in either phase.

\subsection{Two-dimensional kinetic energy and dispersionless states}\label{sec:ekin}

\begin{figure}
\includegraphics[width=0.47\textwidth, trim = 0 155 0 0, clip]{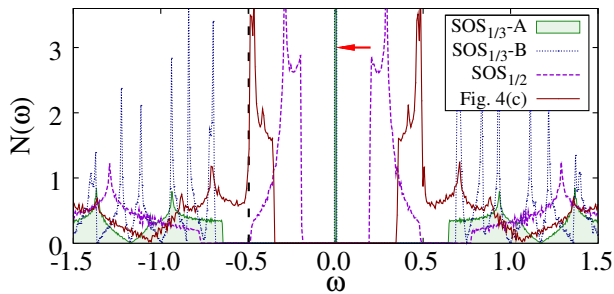}
\caption{(color online) (a) Density of states $N(\omega)$ ($\lambda$=0) for two patterns
of the SOS$_{1/3}$ phase, the phase shown in Fig.~\ref{SOS3}(c), and the
SOS$_{1/2}$ phase.~\cite{Giovannetti:Prl} The arrow indicates the
dispersionless states of the SOS$_{1/3}$ phases.
\label{fig:dos_sos}}
\end{figure}

As discussed above, the effective hoppings with their Berry
phases are not the same in the various (almost) degenerate states of
the SOS$_{1/3}$ phase. It may be tempting to believe that one can find a
local gauge transformation that transforms the sign patters of one
realization into that of another, with a flipped stripe. In this case,
the electronic Hamiltonians would be equivalent and thus have the same
eigenenergies, explaining the degeneracy. However, the density of
states shown in Fig.~\ref{fig:dos_sos} is very different for different
SOS$_{1/3}$ patterns, indicating that the eigenenergies of their
electronic Hamiltonians are in fact quite different.

\begin{figure}
\centering
\subfigure{\includegraphics[width=0.4\textwidth,trim = 0 35 0 0 ,clip]{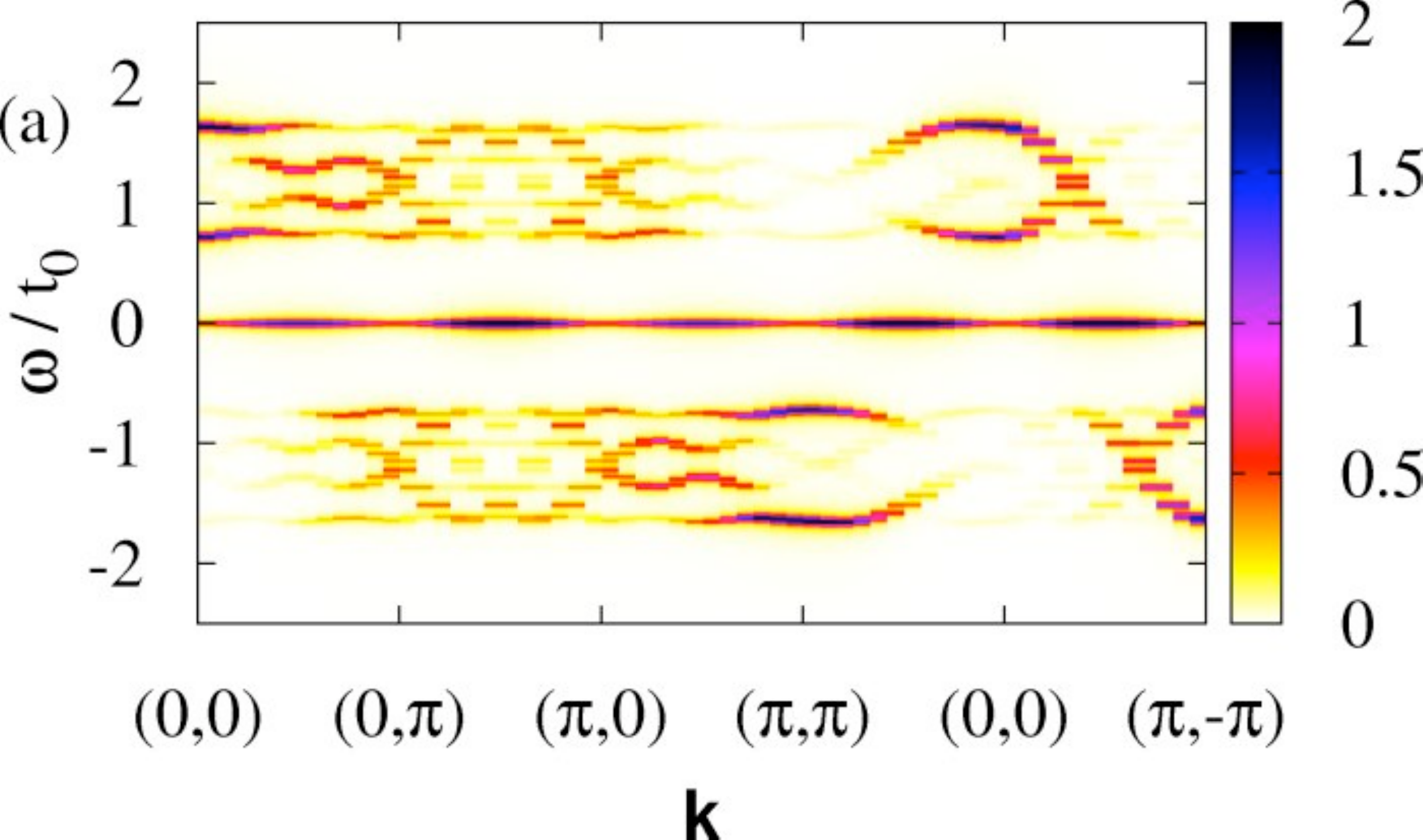}\label{fig:Akw_ref}}
\subfigure{\includegraphics[width=0.4\textwidth,trim = 0 0 0 0 ,clip]{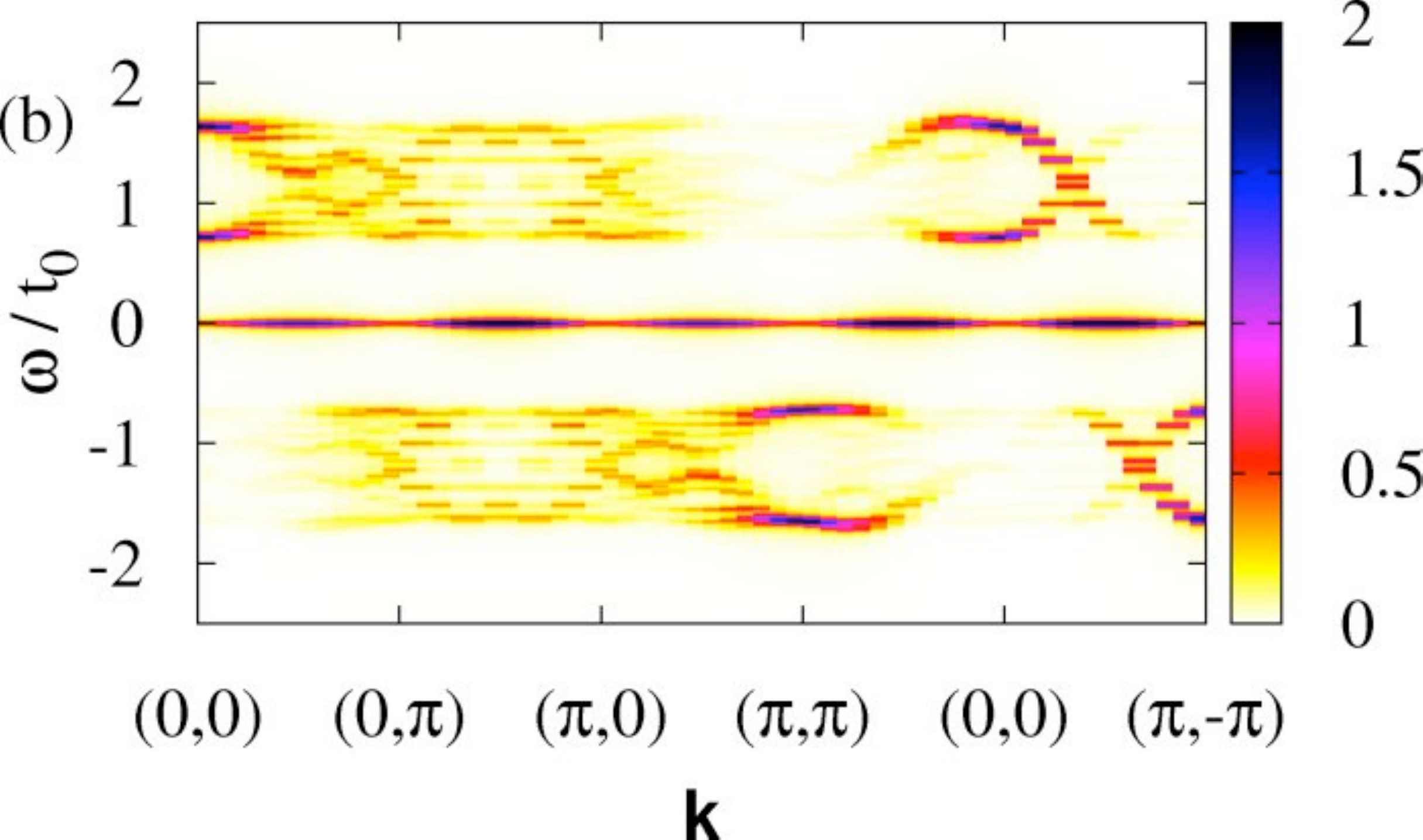}\label{fig:Akw_flip}}
\caption{(color online) Spectral density $A({\bf k},\omega)$ for two patterns
of the SOS$_{1/3}$ phase at $\lambda=0$. The $L_x\times L_y = 24\times24$ lattice is used, and the $E$-AF
domains run in the $(1,1)$ direction, from $(0,0)$ to $(24,24)$. Periodic
boundary conditions are employed, and peaks were broadened with a Lorentzian with a width
$\delta = 0.02t_0$.
\label{fig:Akw}} 
\end{figure}

That the different SOS$_{1/3}$ configurations have different
electronic Hamiltonians can also be inferred from the one-particle spectral density $A({\bf
  k},\omega)$ 
shown in
Fig.~\ref{fig:Akw}. One aspect to note is that the occupied states do
not show any (quasi-) 1D character, and the states are dispersive both along the stripes
[$(0,0)$ - $(\pi,\pi)$] and perpendicular to them [$(0,0)$
- $(\pi,-\pi)$]. In fact, the kinetic energy between the
stripes is necessary for the SOS$_{1/3}$ phase, as the orthogonal
spin arrangements are stabilized by the competition between the FM DE and AF SE
processes. Electrons can not hop directly along the stripes, as NN
spins are AF ordered, but they move along this direction via neighboring
stripes, leading to a somewhat weaker dispersion than perpendicular to
the stripes.

\begin{figure}
\includegraphics[width=0.4\textwidth]{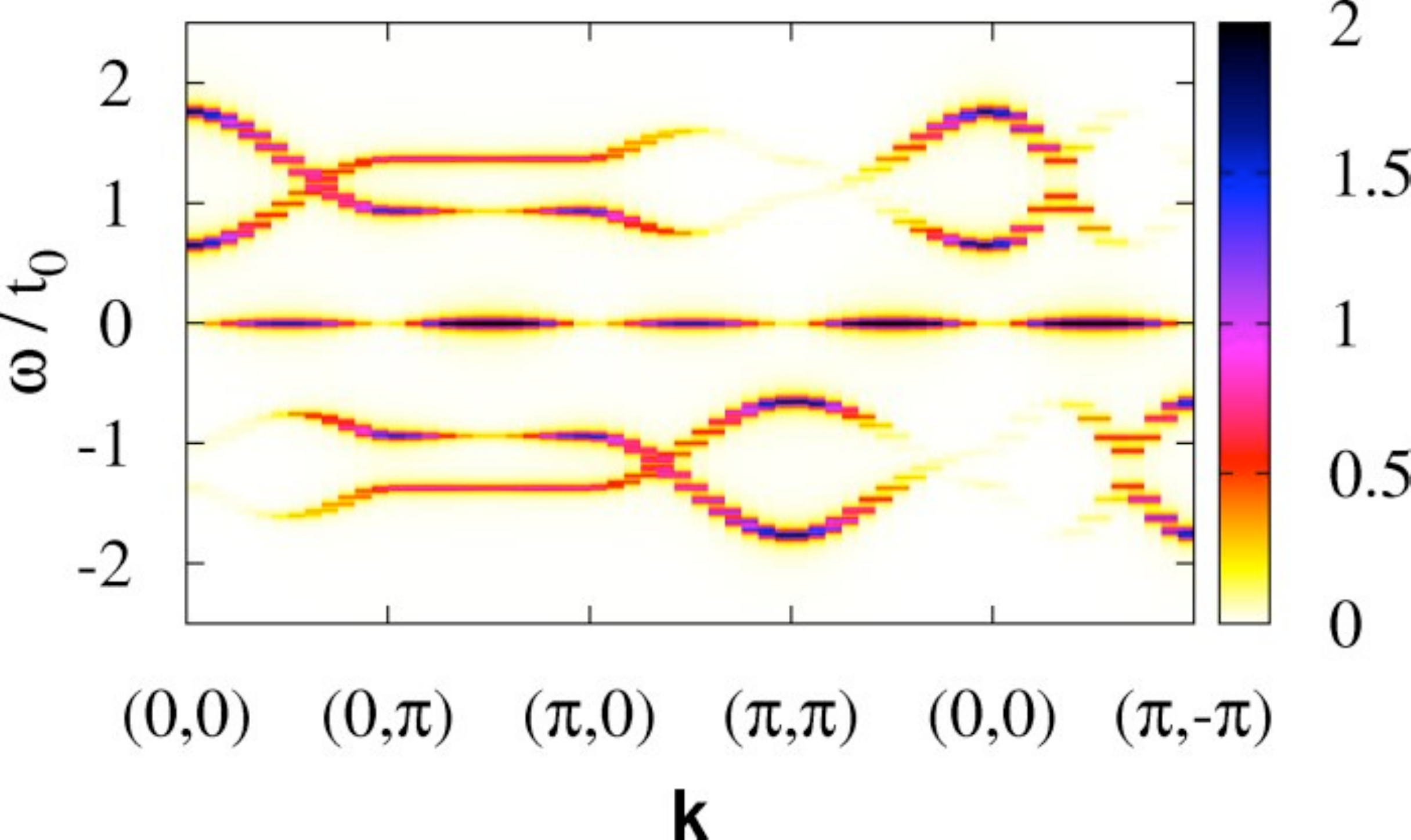}
\caption{(Color online) Spectral density $A({\bf k},\omega)$ for  an
  artificial reference model where all hoppings are replaced by their
  absolute values; this phase has the same energy as the SOS$_{1/3}$
  phase. Parameters are as in Fig.~\ref{fig:Akw}.\label{fig:Ak_abs}} 
\end{figure}

The spectral density of the SOS$_{1/3}$ phase in Fig.~\ref{fig:Akw}
can be compared to that of a toy model where all hoppings are replaced
by their absolute values. As can be seen in Fig.~\ref{fig:Ak_abs},
the spectral density then consists of six coherent bands  (the
dispersionless band at $\omega=0$  is doubly degenerate) corresponding
to six states formed by the two orbitals of three sites within each
arrow. The results for the full model in Fig.~\ref{fig:Akw}
shows some 
remnants of these bands, especially around $(0,0)$ and $(\pi,\pi)$. At
other momenta, however, the bands split or the weight even appears
incoherent. This reflects the additional modulation of the hoppings by
the complex Berry phase, which can lead to superstructures and to 
effective disorder. As one can infer from the density of states, see
Fig.~\ref{fig:dos_sos}, spectral weight is not only shifted to
different momenta in different SOS$_{1/3}$ configurations, but also
transferred between different energies. 

\begin{figure}
\centering
\includegraphics[width=0.35\textwidth]{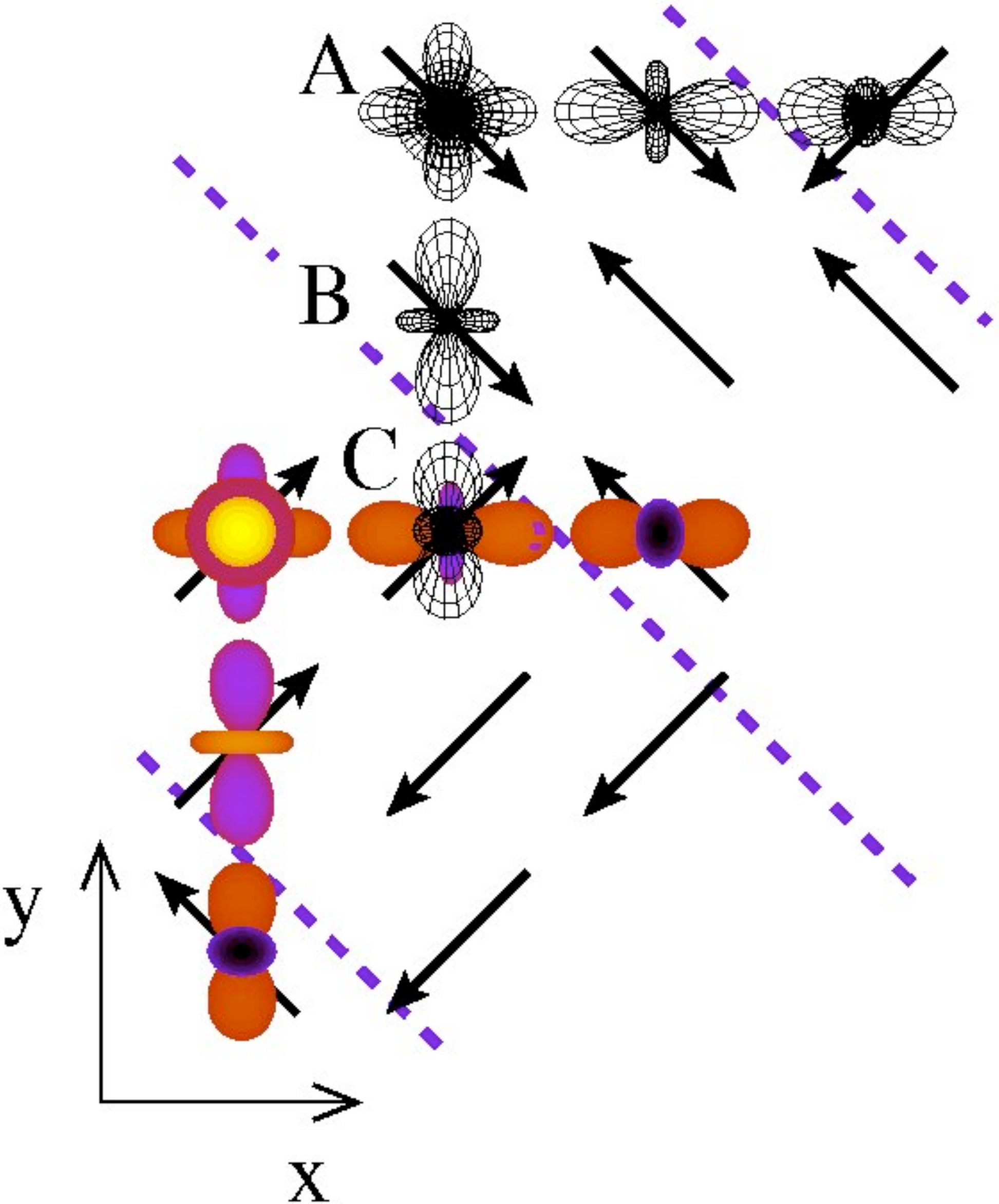}
\caption{(Color online) Two orbital building blocks of the
  SOS$_{1/3}$ phase. Black arrows give the
  SOS$_{1/3}$ spin pattern and dashed lines indicate the stripes.
  Orbitals shown with shading/lines illustrate the unit cells
  corresponding to two arrows in adjacent stripes. Site A is a
  `center' site with occupancy in both the $x^2-y^2$ and the
  $3z^2-r^2$ orbitals, site B is a `wing' site, only the more highly
  occupied $3y^2-r^2$ orbital is shown. Site C is also a `wing' site,
  but here both orbitals are shown, the planar  $y^2-z^2$ orbital is
  drawn with lines and the directional $3x^2-r^2$ with shading. An electron in the planar
  $y^2-z^2$ orbital can \emph{only} hop along $+y$ to site B of the adjacent
  stripe: It cannot hop along the $x$-direction due to its orbital
  symmetry,\cite{Dagotto:Prp} and it cannot hop along the $-y$ bond due to the AF
  spins. \label{fig:con_stripes} }
\end{figure}

The dimensional reduction unveiled here  
appears linked to the dispersionless 
edge-states of the $E$-AF phase,~\cite{Hotta:Prl,Hotta:Rpp} which are
unoccupied in the SOS$_{1/n}$ phases [see the delta peak in the density of states,
Fig.~\ref{fig:dos_sos}, and the dispersionless states in $A({\bf
  k},\omega)$ in Fig.~\ref{fig:Akw}]. While the one-particle energies
of the occupied bands change, the dispersionless states are unaffected
by flipping a stripe. The ``SOS$_{1/2}$''
phase,~\cite{Giovannetti:Prl} and the modified phase of
Fig.~\ref{SOS3}(c) do not have such dispersionless states and do also not
have degenerate states, and we will thus analyze the dispersionless
states and their impact on the SOS$_{1/3}$ phase.

In finite $E$-AF clusters, the 
dispersionless states are $z^2-x^2$ and $y^2-z^2$ orbitals
localized on sites along the edge. These $z^2-x^2$ ($y^2-z^2$) orbitals are localized
because their orbital symmetry only permits them to hop to a single
site, their neighbor in the $x$-($y$-)direction, but this bond is AF, and
hopping is thus suppressed by the magnetic order. In the bulk, the
bond in the opposite direction would be FM, and the orbital can thus
delocalize, but this bond is missing at the cluster edge. If one
considers a single stripe of the SOS$_{1/3}$ phase as an 
isolated $E$-AF domain, then the electrons can only delocalize within
(i) both orbitals of the central site and (ii) the directional
orbitals pointing towards it on the two wings [these orbitals are shown
in Fig.~\ref{CO3}]. The orthogonal (planar)
orbitals on the wings do not hybridize with any other orbital of the
stripe and are the localized edge states, which are empty at $x=1/3$. 

As discussed above in Sec.~\ref{sec:spinorb}, however, hopping
\emph{between} the stripes is not suppressed and the arrows on
adjacent stripes are thus coupled. As illustrated in
Fig.~\ref{fig:con_stripes}, electrons in the planar orbitals can hop to the
directional orbital on the wing site of an adjacent stripe. This leads
to the subsystem consisting of the six orbitals drawn with black lines in
Fig.~\ref{fig:con_stripes}. After taking into account reflection
symmetry with respect to the central site, the even and odd subspace
each yield a $3\times 3$ matrix 
\begin{equation}\label{eq:arrow}
H_{e/o} = \left(\begin{matrix}
0 & -t_{e/o} & 0\\ 
-t_{e/o} &  0 & -\frac{\sqrt{3}t_0}{2\sqrt{2}} \textrm{e}^{i\phi} \\
0& -\frac{\sqrt{3}t_0}{2\sqrt{2}} \textrm{e}^{-i\phi} &   0
\end{matrix}\right). 
\end{equation}
where subscripts $e$/$o$ denote the even and odd subspace, $t_{e,o}$
is given by $\frac{t_0}{2}$ for the even and $\frac{t_0\sqrt{3}}{2}$
for the odd case,  and $\phi$ denotes 
the complex Berry phase. It can be easily seen that such a matrix
always has one eigenvalue  $\epsilon=0$ and that the corresponding eigenvector
only lives on the first and third sites. Here, this implies that the
directional orbital on the wing sites has no overlap with dispersionless states.  
The only process connecting the building blocks described by
Eq.~(\ref{eq:arrow}) is the hopping
$-\frac{t_0}{2\sqrt{2}} \textrm{e}^{-i\phi}$ 
between two directional orbitals sitting on wing sites belonging to
adjacent stripes, e.g., between the $3y^2-r^2$ orbital on site B and
the $3x^2-r^2$ orbital on site C in Fig.~\ref{fig:con_stripes}. Since it only acts on the directional
orbitals, which are not involved in the $\epsilon=0$ states, the latter
are not affected and remain dispersionless in the SOS$_{1/3}$
phase. In the ``SOS$_{1/2}$'' phase,~\cite{Giovannetti:Prl} the $E$-phase domains are too narrow to
support dispersionless states and in the modified phase of
Fig.~\ref{SOS3}(c), where the ``arrows'' of the $E$-phase point in
opposite directions in adjacent domains, the different orbital pattern
shown in Fig.~\ref{CO3_wrong} also prevents similar dispersionless states. 

Such a  protection of a large ground-state
degeneracy by dispersionless states is reminiscent of spin ice, where
flat bands enforce the ice rules.~\cite{PhysRevLett.95.217201}
The notion that the protection of a degeneracy is caused by the dispersionless
states at $\omega =0 $
can be motivated by observing that the lack of dispersion implies that
the states are completely localized and do not communicate with other,
dispersive, states. 
The energy of the occupied subband centered around $\omega = -\epsilon
\approx -1.2 t_0$ can then only be affected by processes involving the symmetric
unoccupied states at $\omega = +\epsilon$. 
Treating the hybridization between these two subbands
in perturbation theory, valid for
large energy differences $2\epsilon > \tfrac{t_0}{2\sqrt{2}}$, the
second order contribution can be shown to be 
independent of the hopping's complex phase and the third order
contribution drops out entirely. 
If hybridization sensitive to the Berry phase then occurs almost
exclusively 
within the occupied subband, the individual one-particle energies seen in the
density of states can be changed, but the total energy cannot, because all
energy gained by one state is lost by another. 

\section{Discussion and conclusions} \label{sec:concl}

The $E$-AF phase, which provides the stripe building blocks, 
displays FE polarization due to 
exchange-striction.~\cite{Sergienko:Prl,Picozzi:Prl} 
Additionally, the many bonds
with non-collinear spins leads to a sizable Dzyaloshinskii-Moriya
interaction. We found that the FE polarization caused by the 
latter is larger in some configurations and smaller in others. This causes an additional energy
difference, but as long as it is small
compared to the energy separating the SOS$_{1/3}$ ground-state
manifold from excited states, this should not qualitatively alter the
physical behavior that we described here. On the
other hand, the different FE properties might provide a handle to
manipulate the stripes. 
It should finally be noted that
the multiferroic properties of a large collection of nearly degenerate
states differing by the FE polarization has not been investigated thus
far and may be highly non-trivial due to potential interference effects.

To conclude, we have studied a model Hamiltonian appropriate for
narrow-band manganites at small Jahn-Teller coupling using unbiased numerical techniques. We
found that the spins of the 2D system spontaneously undergo dimensional reduction into 1D
stripes for dopings $1/3$, $1/4$, $\dots$ Adjacent stripes
always have spins at a 90$^\circ$ angle, but the spins of stripes at larger
distances are not correlated. The electronic kinetic energy, on the
other hand, remains fully 2D. This indicates that the manganite 
results described here induce the spins to behave in a way analogous to
the (orbital pseudo-) spins in the so-called compass
model, which is also used to  describe protected qbits. However, in
contrast to the compass model, our model Hamiltonian does not commute
with the corresponding symmetry operators, and the effect is thus an
\emph{emergent} property of the groundstate manifold.

\begin{acknowledgments}
 
S.L., C.\c{S}., and E.D. were supported by the U.S. Department of Energy, 
Office of Basic Energy Sciences, Materials Sciences and Engineering Division. 
S.D. was supported by the 973 Projects of China (Grant No. 2011CB922101) and the NSFC (Grant No. 11004027).
M.D. acknowledges support by the Deutsche Forschungsgemeinschaft under the
Emmy-Noether program and  helpful discussions with Z. Nussinov.
\end{acknowledgments}


\end{document}